\documentclass[seceq]{ptptex}%
\usepackage{amssymb}
\usepackage{enumerate}
\usepackage{graphicx}
\usepackage{amsmath}
\usepackage{amsfonts}%
\newcommand{\be}{\begin{equation}}
\newcommand{\ee}{\end{equation}}
\newcommand{\bea}{\begin{eqnarray}}
\newcommand{\eea}{\end{eqnarray}}

\inst{{National Superconducting Cyclotron Laboratory, Michigan
State University, East Lansing, MI 48824, USA}} \abst{The velocity
dependence of the stopping power of swift protons and deuterons in
low energy collisions with hydrogen and helium gas targets is
investigated with the numerical solution  of the time-dependent
Schr\"odinger coupled-channels equations using molecular orbital
wavefunctions. At low projectile energies the stopping is mainly
due to nuclear stopping, charge exchange of the electron, and
excitation of the lowest levels in the target. The second and
third mechanisms dominate at $E < 200$ eV. At lower energies it is
also shown that a threshold effect is responsible for a quick drop
of the energy loss. This investigation sheds more light on the
long standing electron screening problem in fusion reactions of
astrophysical interest.}
\begin{document}

\title{The stopping of low energy ions in reactions of astrophysical interest}
\author{Carlos A. \textsc{Bertulani}}
\maketitle

Nuclear fusion reactions proceed in stars at extremely low energies, e.g., of
the order of $10$ keV in our sun \cite{Cla68,RR88}. At such low energies it is
extremely difficult to measure the cross sections for charged particles at
laboratory conditions due to the large Coulomb barrier. One often uses a
theoretical model to extrapolate the experimental data to the low-energy
region. Such extrapolations are sometimes far from reliable, due to unknown
features of the low-energy region. E.g., there might exist unknown resonances
along the extrapolation, or even some simple effect which one was not aware of
\ before. One of these effects is the laboratory atomic screening of fusion
reactions \cite{ALR87,SR95}. It is well known that the laboratory measurements
of low energy fusion reactions are strongly influenced by the presence of the
atomic electrons. This effect has to be corrected for in order to relate the
fusion cross sections measured in the laboratory with those at the stellar
environment. One has observed experimentally a large discrepancy between the
experimental data and the best models available to treat the screening effect
by the electrons in the target nuclei \cite{Ro01}. The screening effect arises
because as the projectile nucleus penetrates the electronic cloud of the
target the electrons become more bound and the projectile energy increases by
energy conservation. Since fusion cross sections increase strongly with the
projectile's energy, this tiny amount of energy gain (of order of 10-100 eV)
leads to a large effect on the measured cross sections. However, in order to
explain the experimental data it is necessary an extra-amount of energy about
twice the expected theoretical value \cite{Ro01}.

In order to extract the fusion cross sections from experiment one needs to
correct for the energy loss in the target to assign the correct projectile
energy value for the reaction. The authors in refs. \cite{LSBR96,BFMH96} have
shown that a possible solution to the long standing discrepancy between theory
and experiment for the reaction $^{3}$He(d$,$ p)$^{4}$He could be obtained if
the projectile energy loss by electronic excitations and charge exchange with
the target atoms would be smaller than previously assumed in the experimental
data analysis. There have been indeed a few experiments in which evidences of
smaller than expected electronic stopping power were reported (see, e.g. ref.
\cite{GS91}). Other reactions of astrophysical interest (e.g., those listed in
by Rolfs and collaborators \cite{ALR87,SR95}) should also be corrected for
this effect. Whereas at higher energies the stopping is mainly due to the
ionization of the target electrons, at the astrophysical energies it is mainly
due to excitations of the lowest levels, charge-exchange between the target
and the projectile, and the nuclear stopping power.

The static two-center p+H system has been solved by Edward Teller
in 1930 \cite{Te30}. He showed that as the distance between the
protons decreases the hydrogen orbitals split into two or more
orbitals, depending on its degeneracy in the two-center system.
Analogous problems are well known in quantum systems. For example,
take two identical potential wells at a certain distance. For
large distances the states in one well are degenerated with the
states in the other potential well. As they approach this
degeneracy is lifted due to the influence of barrier tunnelling.
Thus, the lowest energy state of hydrogen, $1$s, splits into the
1s$\sigma$ and the 2p$\sigma$ states as the protons approach each
other. The 1s$\sigma$ state is space symmetrical, while the
2p$\sigma$ state is antisymmetric. As the proton separation
distance decreases their respective energies decrease. At
$R\simeq1$ \AA \ the energy of the 2p$\sigma$ state starts to
increase again, while the energy of the 1s$\sigma$ state continues
to decrease. For proton distances much smaller than $1$ \AA \ the
1s$\sigma$ and the 2p$\sigma$ energies correspond to those of the
first and second states of the He atom, respectively \cite{Te30}.

The full time-dependent wavefunction for the system can be expanded in terms
of two-center states, $\phi_{n}(t)$, governed by the Schr\"{o}dinger's
equation. \ For the p+H system and at very low proton energies $\left(
E_{p}\lesssim1\text{ keV}\right)  $ it is fair to assume that only the
low-lying states are involved in the electronic dynamics. Only at proton
energies of order of 25 keV the proton velocity is comparable to the electron
velocity, v$_{e}\simeq\alpha c$. Thus, the evolution of the system is almost
adiabatic at $E_{p}\lesssim10$ keV. The higher states require too much
excitation energy and belong to different degeneracy multiplets. The initial
electronic wavefunction is a clear superposition of 1s$\sigma$ and 2p$\sigma$
two-center states. One thus expects that only these states are relevant for
the calculation. In fact, at these energies the population of the 2p atomic
state in charge exchange is much less than the population of the 1s atomic
state. These assumptions are well supported by the calculations of Grande and
Schwietz \cite{GS93}, who have used a dynamical approach based on
target-centered wavefunctions. In their approach one has to include a great
amount of target-centered states in order to represent well the strong
distortion of the wavefunction as the projectile closes in the target. We also
have assumed that the proton follows a classical trajectory determined by an
impact parameter $b$.

If one includes only the two lowest energy molecular states in the p+H system,
Schr\"{o}dinger's equation becomes \cite{BD00}%
\begin{equation}
i\hbar\frac{d}{dt}\left(
\begin{array}
[c]{c}%
a_{+}\\
a_{-}%
\end{array}
\right)  =\left(
\begin{array}
[c]{cc}%
V_{+}+E_{0} & iW\\
iW & V_{-}+E_{0}%
\end{array}
\right)  \left(
\begin{array}
[c]{c}%
a_{+}\\
a_{-}%
\end{array}
\right)  \;, \label{coupled}%
\end{equation}
where the indices $+$ and $-$ refer to the $1$s$\sigma$ and $2$p$\sigma$
states, respectively, $E_{0}=-13.6$ eV, $V_{\pm}\left(  t\right)  =E_{\pm
}(t)-E_{0}$, and and $W\left(  t\right)  $ is the residual potential
\cite{BD00}. We use the formalism of Teller \cite{Te30} to calculate the
wavefunctions $\Psi_{\pm}\left(  R\right)  $ at different inter-proton
distances, $R(t)$, corresponding to a particular time $t$. The static
Schr\"{o}dinger equation is solved in elliptical coordinates. This yields two
coupled differential equations which can be solved by expanding the solutions
in Taylor series. A set of recurrence relations is obtained for the expansion
coefficients when the boundary conditions are used. The energies
$E_{1\mathrm{s}\sigma}\left(  R\right)  $ and $E_{2\mathrm{p}\sigma}\left(
R\right)  $ are obtained by adjusting the constant which separates the two
coupled equations \cite{BD00} to its correct matching value.

It was further shown in ref. \cite{BD00} that the potentials $V_{\pm}\left(
t\right)  $ extend much farther out than $W\left(  t\right)  .$ Moreover, as
$E_{p}$ decreases the potential $W$ decreases faster than the projectile's
velocity, $v_{p}$. At $E_{p}\simeq100$ eV the potential $W$ loses its
relevance as compared to $V_{\pm}$, which have no dependence on $v_{p}$. When
one sets $W=0$ in eq. $\left(  \mathrm{\ref{coupled}}\right)  $, the equations
decouple and it is straightforward to show that the exchange probability is
given by%
\begin{equation}
P_{exch}=\frac{1}{2}+\frac{1}{2}\cos\left\{  \frac{1}{\hbar}\int_{-\infty
}^{\infty}\left[  E_{-}\left(  t\right)  -E_{+}\left(  t\right)  \right]
dt\right\}  \;. \label{approx}%
\end{equation}
At $E_{p}=10$ keV there is an appreciable difference between the full
calculation and the approximation (\ref{approx}). But, for $E_{p}=100$ eV the
results are practically equal, except for very small impact parameters at
which the potential $W$ still has an effect.

The exchange probability is not constant at small impact parameters, but
oscillates wildly around 0.5, specially for low projectile energies. One might
naively assume that because the collision is almost adiabatic, the system
loses memory of to which nucleus the electron is bound after the collision.
Thus, for small impact parameters one would expect a 50\% probability of
finding the electron in one of the nuclei at $t=\infty$. However, this is not
what happens. From eq. (\ref{approx}) we see that minima of the probability
occur for impact parameters satisfying the relation $\int_{-\infty}^{\infty
}\left[  E_{-}\left(  t\right)  -E_{+}\left(  t\right)  \right]  dt=2\pi
\hbar\left(  n+1/2\right)  ,\;\ \ n=0,1,2,...,N.$

This relation looks familiar, of course. It simply states that the
interference between the 1s$\sigma$ and the 2p$\sigma$ states
induces oscillations in the exchange probability. The electron
tunnels back and forth between the projectile and the target
during the ingoing and the outgoing part of the trajectory. When
the interaction time is an exact multiple of the oscillation time,
a minimum in the exchange probability occurs. The average
probability over the smaller impact parameters is indeed 0.5. As
the impact parameter decreases from infinity, the first maximum in
the exchange probability indicates the beginning of the region of
strong exchange probability. At low proton energies this starts at
$b\simeq3$ \AA . The size of the hydrogen atom is about 0.5 \AA \
and thus the electron travels in a forbidden region (tunnels) of
about 2 \AA \ from the target to the projectile. This is possible
because of the strong interference between the 1s$\sigma$ and the
2p$\sigma$ states, which for some trajectories satisfy the quantum
relation above.

To obtain the stopping power we need the total cross section for charge
exchange, $\sigma=2\pi\int P_{exch}bdb$. For $E_{p}\longrightarrow0$, the
charge exchange cross section becomes the constant value $\sigma\left(
E_{p}=0\right)  =37.88$ $\times10^{-16}$ cm$^{2}$. This happens because, when
$E_{p}\longrightarrow0$ and as the projectile nears the targets, the
increasing electron binding in the two-center system acts as a push in the
relative motion energy to compensate for energy conservation. The average
result is that the cross section for charge exchange becomes approximately
constant for projectile energies of tens of eV and below.

\begin{figure}[ptb]
\begin{center}
\includegraphics[height=3.in, width=2.9in]{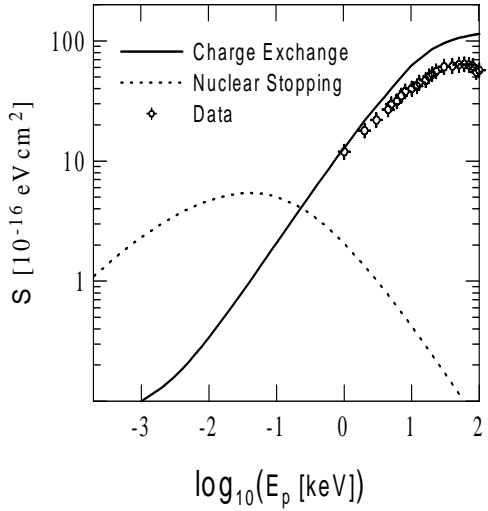}\label{fig}
\end{center}
\caption{The stopping cross section of protons on H-targets. The dotted line
in gives the energy transfer by means of nuclear stopping, while the solid
line are our results for the charge-exchange stopping mechanism. The data
points are from the tabulation of Andersen and Ziegler \cite{AZ77}. }%
\label{fig4}%
\end{figure}

In figure \ref{fig4} we show the stopping cross section of the proton. The
stopping cross section is defined as $S=\sum_{i}\Delta E_{i}\;\sigma_{i}$ ,
where $\Delta E_{i}$ is the energy loss of the projectile in a process denoted
by $i$. The stopping power, $S_{P}=dE/dx$, the energy loss per unit length of
the target material, is related to the stopping cross section by $S=S_{P}/N$,
where $N$ is the atomic density of the material. In the charge exchange
mechanism the electron is transferred to the ground state of the projectile
and the energy transfer is given by $\Delta E=m_{e}v_{p}^{2}/2$, where $v_{p}$
is the projectile velocity. Assuming that there are a few free electrons in
the material (e.g., in a hydrogen gas) only one more stopping mechanism at
very low energies should be considered: the nuclear stopping power. This is
simply the elastic scattering of the projectile off the target nuclei. The
projectile energy is partially transferred to the recoil energy of the target
atom. The stopping cross section for this mechanism has been extensively
studied in ref. \cite{ZBL85}. \ The nuclear stopping includes the effect of
the electron screening of the nuclear charges.

The dotted line in figure \ref{fig4} gives the energy transfer by means of
nuclear stopping, while the solid line are our results for the charge-exchange
stopping mechanism. The data points are from the tabulation of Andersen and
Ziegler \cite{AZ77}. We see that the nuclear stopping dominates at the lowest
energies, while the charge-exchange stopping is larger for proton energies
greater than 200 eV. Since we neglect the difference between molecular and
atomic hydrogen targets, there is a limitation to compare our results with the
experimental data. But, the order of magnitude agreement is very good in view
of our simplifying assumptions. We do not consider the change of the charge
state of the protons as they penetrate the target material. The exchange
mechanism transforms the protons into H atoms. These again interact with the
target atoms. The can loose their electron again by transfer to the 1s state
of the target \cite{GS93}.

The best fit to our calculation for the stopping power for proton
energies in the range 100 eV - 1 keV yields $S\sim v_{p}^{1.35}$.
This contrasts with the extrapolation $S\sim v_{p}$, based on the
Andersen-Ziegler tables.

Let us now consider the systems p+$^{4}$He and d+$^{3}$He. The situation is
more complicated because of the electron-electron interaction. The atomic
wavefunctions, $\phi_{\mu}=\sum_{j}c_{j\mu}\phi_{j}^{Slat}$, are constructed
as a linear combination of Slater-type orbitals (STO) \cite{Lev00} of the form
$\phi_{n}^{Slat}=Nr^{n-1}\exp\left(  -\zeta r\right)  Y_{lm}\left(
\widehat{\mathbf{r}}\right)  $, where the Slater coefficients $n$ and $\zeta$
are chosen to best approximate the exact atomic wavefunctions (see, e.g. ref.
\cite{Lev00}). The molecular orbital wavefunctions for the $\mathrm{p}%
+\mathrm{He}$ system, are obtained with the $\phi_{\mu}$'s chosen so that half
of the STO's are centered on the proton $(A)$ and the other half are centered
on the helium nucleus $(B)$. The total wavefunction for the two-electron
system is finally written as a Slater determinant of the molecular orbital
wavefunctions. Configuration-interaction with double excitation configurations
were included in the calculation \cite{bert04}, with the coefficients $n$ and
the Slater parameters $\zeta$\ chosen in a variational method to obtain the
lowest energy states of the system.

Using these conditions and variation method, one obtains the Hatree-Fock
equations for the electronic states. Solving the Hartree-Fock equations one
obtains the coefficients $c_{ij}$ which give the proper linear combination of
atomic orbitals to form the molecular orbital.

\begin{figure}[ptb]
\begin{center}
\includegraphics[
height=3.4in, width=2.9in
]{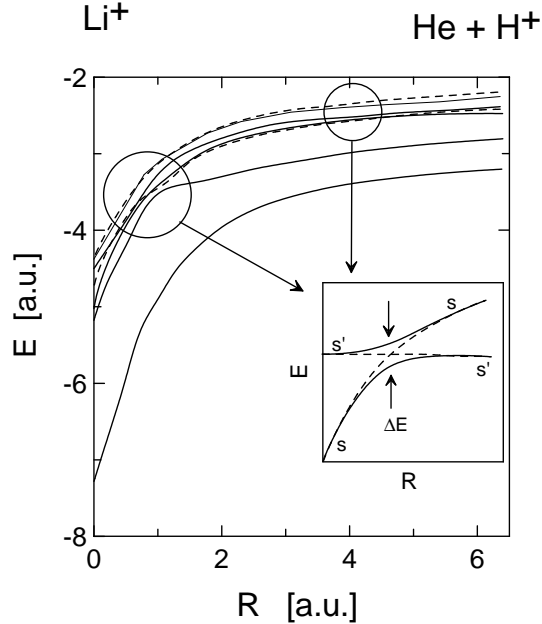}
\end{center}
\caption{Adiabatic energies (1 a.u. of energy = 27.2 eV, 1 a.u. of length =
0.53 \AA ) for the electronic orbitals for the (H-He)$^{+}$ system as a
function of the internuclear separation. As the atoms approach each other
slowly curves of same symmetry repel each other. A transition between states s
and s' can occur in a slow collision. In a fast collision a diabatic
transition, with the states crossing each other, will occur. This is shown in
the inset.}%
\label{f1}%
\end{figure}

In figure \ref{f1} we show the intersection points of the states with same
symmetry in the H$^{+}$+He system. In a fast collision these states would
cross (diabatic collisions), whereas in collisions at very low energies
(adiabatic collisions) they obey the von Neumann-Wigner non-crossing rule.

In the dynamical case the full time-dependent wavefunction for the system can
be expanded in terms of two-center states,\ with expansion coefficients
$a_{n}\left(  t\right)  $. The dynamical evolution of the H+He system is
calculated using the same approach as described in ref. \cite{BD00}. The
solutions are obtained starting from initial internuclear distance of 15 a.u.
for the incoming trajectory and stopped at the same value for the outgoing
trajectory. The probability for the capture in the proton is obtained by a
projection of the final wavefunction into the wavefunctions of \ the $1s$,
$2s$ and and $2p$ states of the hydrogen atom.

A similar situation occurs for $\mathrm{p}+\mathrm{He}$ collisions
for the electron capture probability by the proton at a few keV
bombarding energy. There will be oscillations due to the electron
exchange between the ground state of the hydrogen and the first
excited state in He (1s2s). But, in contrast to the H$^{+}$H
system, the oscillations are strongly damped. Following the work
of Lichten \cite{Lich63}\ we interpret this damping effect as due
to the interference between the participant states and a band of
states of average energy $\left\langle E_{a}\right\rangle $\ and
width 2$\Gamma$, as seen in figure \ref{f1}. The important regions
where the diabatic level cross occurs is shown in figure \ref{f1}
inside the encircled areas. The damping mechanism is best
understood using the Landau-Zener theory for level crossing. At
the crossing there is a particular probability ($1-P$)
of an adiabatic transition where $P$ is given by the Landau-Zener formula%
\begin{equation}
P_{exch}=\exp\left[  \frac{2\pi H_{ss^{\prime}}^{2}}{v\left(  d/dR\right)
\left(  E_{s}-E_{s^{\prime}}\right)  }\right]  \label{lz}%
\end{equation}
where $v$ is the collision velocity and $H_{ss^{\prime}}$ is the off-diagonal
matrix element connecting states $s$ and $s\prime$. The oscillatory behavior
of the exchange probability is due to the many level transitions at the
crossing. The interference with the neighboring states introduces a damping in
the charge exchange probability, i.e.%
\[
P_{exch}\left(  b,t\right)  \simeq\cos^{2}\left(  \frac{\left\langle
E_{a}\right\rangle b}{v}\right)  \exp\left[  -\frac{2\pi\Gamma^{2}%
b}{v\left\langle E_{a}\right\rangle }\right]  ,
\]
where $\left\langle E_{a}\right\rangle \simeq1.$ a.u. is the average
separation energy between the 0$\Sigma$ level and the bunch of higher energy
levels shown in figure \ref{f1}. The exponential damping factor agrees with
the numerical calculations if one uses $\Gamma\simeq5$ eV, which agrees with
the energy interval of the band of states shown in figure \ref{f1}.

At very low energies the only possibility that the electron is captured by the
proton is if there is a transition 1s$^{2}$($^{1}$S$_{0}$) $\longrightarrow$
1s2s($^{3}$S) in the helium target. Only in this case the energy of one of the
electrons in helium roughly matches the electronic energy of the ground state
in H. This resonant transfer effect is responsible for the large capture cross
sections. When this transition is not possible the electrons prefer to stay in
the helium target, as the energy of the whole system is lowest in this case.
Another possible mechanism for the stopping is the excitation of the helium
atom by the transition 1s$^{2}$($^{1}$S$_{0}$) $\longrightarrow$ 1s2s($^{3}%
$S). Thus, there must be a direct relationship between the energy transfer to
the transition 1s$^{2}$($^{1}$S$_{0}$) $\longrightarrow$ 1s2s($^{3}$S) and the
minimum projectile energy which enables electronic changes. Ref. \cite{For00}
reported for the first time this effect, named by threshold energy, which can
be understood as follows. The momentum transfer in the projectile-target
collision, $\Delta q$, is related to the energy transfer to the electrons by
$\Delta q=\Delta E/v$, where $v$ is the projectile velocity. In order that
this momentum transfer absorbed by the electron, induces an atomic transition,
it is necessary that $\hbar^{2}\Delta q^{2}/2m_{e}\sim\Delta E$. Solving these
equations for the projectile energy one finds
\begin{equation}
E_{p}^{thres}\sim\frac{m_{p}}{4m_{e}}\Delta E\ . \label{ethres}%
\end{equation}
This is the threshold energy for atomic excitations and/or charge exchange. If
the projectile energy is smaller than this value, no stopping should occur.
The energy for transition 1s$^{2}$($^{1}$S$_{0}$) $\longrightarrow$
1s2s($^{3}$S) in He is $\Delta E=18.7$ eV. Thus, for $\mathrm{p}+\mathrm{He}$
collisions, the threshold energy is $E_{p}^{thres}\sim9$
keV.

\begin{figure}[ptb]
\begin{center}
\includegraphics[
height=3.3in, width=3.in ]{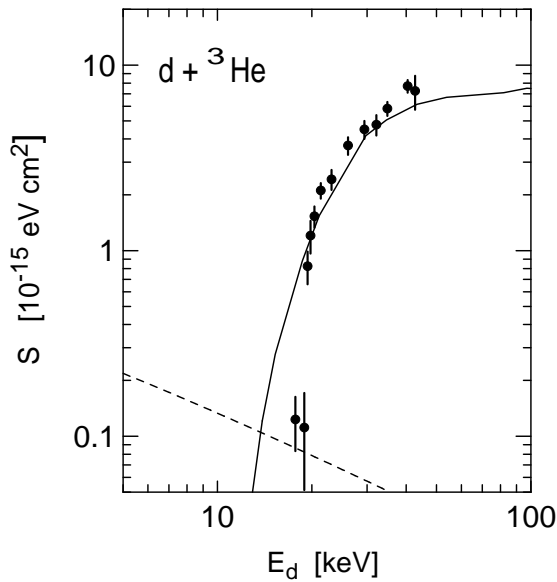}
\end{center}
\caption{Energy loss of deuterons in $^{3}\mathrm{He}$ gas as a function of
deuteron energy. Data are from ref. \cite{For00}. The solid curve is the
calculation for the electronic stopping power, while the dashed curve shows
the nuclear stopping.}%
\label{stopHe3}%
\end{figure}

Figure \ref{stopHe3} shows the energy loss of deuterons in $^{3}\mathrm{He}$
gas as a function of deuteron energy. The data are from ref. \cite{For00}. The
solid curve is the numerical calculation for the electronic stopping power,
while the dashed curve shows the nuclear stopping. As discussed in ref
\ \cite{For00} the threshold deuteron energy in this reaction is of the order
of 18 keV, which agrees with the estimate based on eq. \ref{ethres}. However,
the numerical calculations based on the electronic stopping (solid curve of
fig. \ref{stopHe3}) indicate a lower threshold energy for this system.
Nonetheless, the agreement with the experimental data is very good for
$E_{d}>20$ keV. The threshold effect is one more indication that the
extrapolation $S\sim v$, based on the Andersen-Ziegler tables is not
applicable to very low energies.

The steep rise of the fusion cross sections at astrophysical energies
amplifies all effects leading to a slight modification of the projectile
energy \cite{BBH97}. The results presented here show that the stopping
mechanism does not follow a universal pattern for all systems. The threshold
effect reported in ref. \cite{For00} is indeed responsible for a rapid
decrease of the electronic stopping at low energies. It will occur whenever
the charge-exchange mechanism and the excitation of the first electronic state
in the target involve approximately the same energy. However, the drop of the
electronic stopping is not as sharp as expected from the simple classical
arguments given by eq. \ref{ethres}.

The experiments on astrophysical fusion reactions have shown that the
screening effect is much larger than expected by theory. The solution to this
problem might be indeed the smaller stopping power, due to a steeper slope at
low energies induced, e.g. by the threshold mechanism. This calls for improved
theoretical studies of the energy loss of ions at extremely low energies of
and for their independent experimental verification. The present situation is
highly disturbing because if we cannot explain the laboratory screening
effect, most likely we cannot explain it in stellar environments.


\bigskip

\begin{thebibliography}{99}                                                                                               %


\bibitem {Cla68}D.D. Clayton, \textit{Principles of Stellar Evolution and
Nucleosynthesis}, McGraw-Hill, New York, 1968

\bibitem {RR88}C.. Rolfs and W.S. Rodney, \textit{Cauldrons in the Cosmos},
Chicago Press, Chicago, 1988

\bibitem {ALR87}H.J. Assenbaum, K. Langanke, and C. Rolfs, Z. Phys.
\textbf{A327}, 461 (1987)

\bibitem {SR95}E. Somorjai and C. Rolfs, Nucl. Instum. Meth. \textbf{B99}, 297 (1995)

\bibitem {Ro01}C. Rolfs, Prog. Part. Nucl. Phys. \textbf{46}, 23 (2001)

\bibitem {LSBR96}K. Langanke, T.D. Shoppa, C.A. Barnes and C. Rolfs, Phys.
Lett. \textbf{B369}, 211 (1996)

\bibitem {BFMH96}J.M. Bang, L.S. Ferreira, E. Maglione, and J.M. Hansteen,
Phys. Rev. \textbf{C53}, R18 (1996)

\bibitem {GS91}R. Golser and D. Semrad, Phys. Rev. Lett. \textbf{14}, 1831 (1991)

\bibitem {Te30}E. Teller, Z. Physik, \textbf{61}, 458 (1930)

\bibitem {GS93}P.L. Grande and G. Schiwietz, Phys. Rev. \textbf{A47} (1993)
1119; Phys. Rev.\textbf{\ A58}, 3796 (1998); Nucl. Inst. Meth. \textbf{B153},
1 (1999)

\bibitem {BD00}C.A. Bertulani and D.T. de Paula, Phys. Rev. \textbf{C 62},
045802 (2000)

\bibitem {ZBL85}J.F. Ziegler, J.P. Biersack and U. Littmark,
{\it The stopping and range of ions in matter}, Vol. 1, Pergamon
Press, New York (1985)

\bibitem {AZ77}H. Andersen and J.F. Ziegler, \textit{The stopping and ranges
of ions in matter}, Vol. 3, Pergamon press, New York (1977)

\bibitem {Lev00}I.N. Levine, \textit{Quantum Chemistry}, 5th ed., Prentice
Hall (2000)

\bibitem {bert04}C.A. Bertulani, nucl-th/0401002

\bibitem {Lich63}W. Lichten, Phys. Rev. \textbf{131} (1963) 229

\bibitem {For00}A. Formicola et al., Eur. Phys. J. \textbf{A8}, 443 (2000)

\bibitem {BBH97}A. B. Balantekin, C. A. Bertulani, M. S. Hussein, Nucl. Phys.
\textbf{A627}, 324 (1997)

\end{thebibliography}
\end{document}